\documentclass[11pt]{article}

\usepackage[preprint]{acl}

\usepackage{times}
\usepackage{latexsym}
\usepackage[T1]{fontenc}

\usepackage[utf8]{inputenc}

\usepackage{microtype}

\usepackage{inconsolata}

\usepackage{graphicx}
\usepackage{subcaption}
\usepackage{xcolor}
\usepackage{mdframed}
\usepackage{multirow}
\usepackage{booktabs}
\usepackage[table,xcdraw]{xcolor}
\title{Exploring the Effectiveness of Using LLMs for Automated Assessment of Student Self--Explanations in Programming Education}

\author{
 \textbf{Arun-Balajiee Lekshmi-Narayanan},\\
 \textbf{Mohammad Hassany},\\
 \textbf{Peter Brusilovsky}
\\
University of Pittsburgh,
\\
 \small{
   \textbf{Correspondence:} \href{mailto:arl122@pitt.edu}{arl122@pitt.edu}
 }
}

\begin{document}
\maketitle
\begin{abstract}

Worked examples are step-by-step solutions to problems in a specific domain, offered to students to acquire domain-specific problem-solving skills. The effectiveness of worked examples could be enhanced by combining them with self-explanations, which ask students to explain rather than passively study each problem-solving step. The main challenge of this approach is assessing the correctness of the student's explanations. In the prevailing approach, student explanations are judged by their semantic similarity to an instructor's or domain expert's explanation. Given recent advances in LLM-based automated scoring,  it remains unclear whether semantic similarity methods are still the most effective technique to automatically score textual student responses like essays or code explanations. Comparing these methods also requires quality datasets that offer distinctive features  such as balanced class distributions and domain--specific labeled data for automated scoring tasks. In this paper, we present a rigorous comparison between LLMs and semantic similarity used for automated scoring, framed as a binary classification task. 
Our results 
suggest that LLM approaches (F1 = 0.98, accuracy = 0.96) outperform semantic similarity methods (F1 = 0.72, accuracy = 0.65). 

\end{abstract}

\section{Introduction}

In the domain of programming, worked examples (WE) are offered as ``explained code'' where the lines or fragments of a program (code) solving a problem are explained by experts~\citep{linn1992can,brusilovsky2008}. Summarizing multiple years of research on learning, the ICAP framework~\citep{chi2014icap} argues that active learning is more effective than passive learning. An efficient way to turn learning from examples from passive to active is \emph{self-explanations}, where learners are prompted to explain WE themselves rather than reading expert explanations~\citep{chi1989self,chi1994eliciting}. Currently, this approach is used in both assessment and study contexts. In \emph{an assessment context}, instructors include code explanation questions in home assignments and exams. In \emph{a study context}, students are encouraged to self-explain code-based WE either by adding explanatory comments to WE code~\citep{garces2023engaging} or by writing code explanations to their workbooks~\citep{Vihavainen_2015}.

Evaluating student-written explanations is a hard task for instructors, both in terms of grading and providing corrective feedback. As a result, such questions are rarely assigned in take-home assessments, and when they do appear in study contexts, the evaluations are often neglected entirely. This leaves the learning process incomplete: students need proper feedback on their written explanations so that they do not form dangerous misconceptions~\citep{kerslake2025exploring}. Good automatic assessment systems could help fill this gap.

Recent advances in natural language processing (NLP) have enabled automatic evaluation of student self-explanations by comparing them to expert explanations using semantic similarity by representing the sentences using vector embeddings~\citep{reimers2019sentence, chapagain2022automated,oli2024exploring}.
Further, 
recent LLM-as-a-judge has been explored as a potentially more direct alternative to semantic similarity. For example, ~\cite{liu-etal-2023-predicting,hou2025improve} independently demonstrated 2 different methods to automatically score student essays. 

Within the context of code self--explanations, \cite{pmlr-v257-oli24a} investigated a finetuned open-source LLM with preference optimization to generate feedback for students in a dialog-based intelligent tutoring system. \cite{oli2024can} used prompt engineering to assess LLMs' ability to identify gaps and student misconceptions when comprehending code.
~\cite{denny2024explaining,kerslake2025exploring} had students demonstrate their code comprehension skills by explaining the code of small programs. However, none of these papers compare semantic similarity with LLM--powered automated assessments.


A problem with reliably comparing the success of LLMs in automatically evaluating students' responses with semantic similarity methods is the paucity of quality datasets specific to the task. A recent popular dataset~\cite{selfcode2_dataset} contains 1854 pairs of student and expert explanations, 1794 student explanations are labeled as correct and 60 as incorrect. with a ratio of positive to negative samples of approximately 30:1.  
Traditional oversampling techniques such as SMOTE are commonly used to address class imbalance but require numerical feature representations. 

A growing body of work in educational NLP has explored generative AI for creating synthetic training examples. For example,  ~\cite{lee2024applying} generated negative student responses to guide LLMs' reasoning paths for automated scoring with Chain-of-Thought prompt. ~\cite{filighera2024cheating} generated adversarial examples that imitate students' misconceptions to test and fool LLMs' ability in automated scoring. ~\cite{misgna2024survey} provide a survey of examples where negative samples were generated with LLMs using rubrics created with handcrafted features representative of negative examples. 

To the best of our knowledge this is the first study to use synthetically generated negative examples to balance educational dataset for the specific purpose of comparing semantic similarity and LLM--based approaches to automated assessment of student code explanations. 

Our primary research question is, \emph{Could  LLMs be a new approach that competes with traditional semantic similarity?}. 



We address this question by framing the task as binary classification problem using an existing dataset~\cite{selfcode2_dataset}, where the model predicts whether a given student explanation is correct (positive class) or incorrect (negative class). We evaluate several semantic similarity-based approaches alongside strong LLM baselines under various prompting strategies. Our results show that LLM-based assessment (accuracy = 0.96 and F1 = 0.98)  outperforms semantic similarity baselines (accuracy = 0.65, F1 = 0.72)  on the classification task.

\section{Method}
\subsection{Dataset}
To answer our research question, we used the recently released SelfCode2.0 dataset~\citep{selfcode2_dataset}, which includes explanations produced by 60 students and 2 experts for individual lines of four code examples. In the dataset, each student's explanation is labeled by annotators as correct or incorrect. In total,  the dataset included 3019 pairs of student and expert explanations for the same code lines. For our study, we used a subset of these pairs (1854 out of 3019) in which both a student and an expert explained the line using a single statement. This selection was made to facilitate the exploration of modern semantic-similarity approaches that focus on complete sentences. In these 1854 pairs, 1794 student explanations were labeled as correct (positive examples) and 60 as incorrect negative examples. Our secondary contribution in this work addresses the class imbalance in the dataset by synthetically generated negative examples using LLMs.

\subsection{LLM Based Evaluation}
Our LLM-based approaches for evaluating student explanations are based on treating an LLM-as-a-Judge that directly predicts the \emph{correctness} of student explanations. We prompt the LLM with the context that included the problem statement of the explained WE, the code line to explain along with the line number, and the associated student explanations from the data set (Figure~\ref{fig:prompt_llm_exp_eval} in Appendix). In all prompts, we use the same section headings in capital letters, such as ``PROGRAM DESCRIPTION'', ``LINE NUMBER'' and so on. We used the prompts with GPT-3.5-Turbo-16k for the automated assessment task.

\paragraph{No Definition} We did not define correctness to LLM for evaluation.
\paragraph{Construction} We defined correctness by focusing on the construction-based approach to explain a code line, i.e., the role of the line in solving the programming problem.
\paragraph{Code Behavior Prompt} We defined correctness by focusing on the behavior-based approach to explain a code line, i.e., the result of executing this line. 

\subsection{Semantic Similarity Based Evaluation}

We used three popular versions of semantic modeling approaches:

\noindent\textbf{Deep-Tutor (DT) toolkit} that adopted the algorithm described to evaluate correctness in the DT-GRADE corpus~\citep{banjade2016evaluation} and was used in process of data collection of our dataset~\citep{selfcode2_dataset}.  The method generates similarity scores using Support Vector Regression with a feature set combining SentenceBERT and DeepTutor similarity scores~\citep{rus2013recent}.

\noindent\textbf{RoBERTa~\citep{liu2019roberta}} a version of BERT with appropriate modifications for Next Sentence Prediction tasks, available through sentence-transformers\footnote{\url{https://huggingface.co/sentence-transformers/all-roberta-large-v1}}.

\noindent\textbf{GPT Sentence Embeddings (GPT-SE)}\footnote{\url{https://platform.openai.com/docs/guides/embeddings}} available as an API which creates a vector representation for any form of text, either broken down into sentences, individual tokens, or entire paragraphs using the text-embedding-3-small from OpenAI.

For each approach, we computed semantic similarity scores ranging from 0 to 1 for each pair of student and expert explanations. Deep-Tutor is designed to return this score directly. To obtain a semantic similarity score using RoBERTa or GPT-SE, we generated sentence embeddings for all student and expert explanations. We calculated cosine similarity between the embeddings of student and expert explanations within each pair.

\subsection{Comparing Semantic Similarity and LLM-Based Evaluation on a Balanced Dataset}
To address the class imbalance in the original dataset, we created the balanced dataset with 1794 positive and negative examples. To generate negative examples, we used a small language model (SLM) GPT-OSS 20b model~\cite{openai2025gptoss120bgptoss20bmodel} with the prompt in the Figure~\ref{fig:prompt_generate_negative_examples}. We chose the model because it differs from the one we use to assess student explanations. We selected the lines of code for which student explanations are available in the SelfCode2.0 dataset~\cite{selfcode2_dataset} and prompted the SLM to generate 3 distinct incorrect explanations for every line of code.  We provide more details on prompts and dataset verification in the Appendix~\ref{sec:gen_neg_ex}.



We conducted 5-fold cross-validation on the balanced dataset. The training set in every fold consisted of 1435 positive and negative examples each, and the testing sets consisted of 359 positive and negative examples each. We used the training set to identify the best threshold for the semantic similarity baselines using the optimal F1 scores. In contrast, we used the test sets to compare LLM performance and to assess student explanations using semantic similarity. The LLMs were not trained on the training set, since we were using them in a zero-shot setting.

\section{Results}

As the Table~\ref{tab:comparison_Roberta} shows, the LLM Behavior (F1 = 0.98, Accuracy = 0.96) outperforms Deep Tutor (F1 = 0.69, Accuracy = 0.6) and RoBERTa (F1 = 0.7, Accuracy = 0.61) (see Table~\ref{tab:comparison_Roberta}). The best performance on 5 fold test sets of LLM Behavior \emph{(TP=357,FP=5,FN=1,TN=2)} was also better than Deep Tutor's \emph{(TP=341,FP=317,FN=18,TN=42)}, where TP = True Positives, FP = False Positives, FN = False Negatives and TN = True Negatives. More results are available on Drive~\footnote{\url{https://docs.google.com/spreadsheets/d/1qj5uz1AWwEIkXm8mnlBA-IHeILYulIq4ZofL14pvozs/edit?usp=sharing}} and the Table~\ref{tab:all-tp-fp-fn-tn}. The fewer false negatives for LLM--powered automated scoring suggests that the system evaluates less cases of student explanations that are correct as incorrect. Fewer false negatives makes the system less frustrating for the students to work with when they are assessed incorrect for cases that the students know their answers are correct.

The LLM-Constr. potentially underperformed (for both datasets -- F1 =0.89, Acc = 0.8) because the student explanations in the dataset described the ``syntax and code behavior" and missed out the context of the code within the rest of the program, indicating that the prompts to the LLMs must balance the rubrics to assess students' explanations between the instructors' expectations and students' understanding of a correct explanation for a line of code. The performance of the similarity-based methods being lower than the LLMs could suggest that LLM methods are more robust than similarity-based methods when considering zero-shot performance with educational assessment tasks.



\begin{table}[]
\centering
\tiny
\resizebox{\columnwidth}{!}{%
\begin{tabular}{@{}lll@{}}
\toprule
Model        & F1   & Acc. \\ \midrule
Deep Tutor   & 0.69 & 0.6  \\
RoBERTa      & 0.7  & 0.61 \\
GPT-SE       & 0.72 & 0.65 \\
LLM-Behavior & 0.98 & 0.96 \\
LLM-NoDef.   & 0.98 & 0.96 \\
LLM-Constr.  & 0.89 & 0.8  \\ \bottomrule
\end{tabular}%
}
\caption{Comparing Semantic Similarity and LLM approaches on test sets in 5-Fold CV setup with balanced dataset using artificial data}
\label{tab:comparison_Roberta}
\end{table}

\begin{table}[]
\centering
\resizebox{\columnwidth}{!}{%
\begin{tabular}{@{}lllll@{}}
\toprule
             & TP  & FP  & FN & TN  \\ \midrule
\multicolumn{5}{c}{\textbf{Fold 1}} \\
Deep Tutor   & 341 & 317 & 18 & 42  \\
RoBERTa      & 335 & 284 & 24 & 75  \\
GPT-SE       & 328 & 272 & 31 & 87  \\
LLM Behavior & 355 & 10  & 4  & 2   \\
LLM No Def   & 353 & 11  & 6  & 1   \\
LLM Constr   & 287 & 5   & 72 & 7   \\
\multicolumn{5}{c}{\textbf{Fold 2}} \\
Deep Tutor   & 321 & 250 & 38 & 109 \\
RoBERTa      & 321 & 225 & 38 & 134 \\
GPT-SE       & 313 & 202 & 46 & 157 \\
LLM Behavior & 354 & 9   & 5  & 4   \\
LLM No Def   & 356 & 13  & 3  & 0   \\
LLM Constr   & 290 & 5   & 69 & 8   \\
\multicolumn{5}{c}{\textbf{Fold 3}} \\
Deep Tutor   & 333 & 263 & 26 & 96  \\
RoBERTa      & 320 & 242 & 39 & 117 \\
GPT-SE       & 317 & 214 & 42 & 145 \\
LLM Behavior & 351 & 10  & 8  & 3   \\
LLM No Def   & 350 & 13  & 9  & 0   \\
LLM Constr   & 281 & 5   & 78 & 8   \\
\multicolumn{5}{c}{\textbf{Fold 4}} \\
Deep Tutor   & 322 & 245 & 37 & 113 \\
RoBERTa      & 344 & 256 & 15 & 102 \\
GPT-SE       & 330 & 196 & 29 & 162 \\
LLM Behavior & 357 & 12  & 2  & 3   \\
LLM No Def   & 356 & 15  & 3  & 0   \\
LLM Constr   & 310 & 10  & 49 & 5   \\
\multicolumn{5}{c}{\textbf{Fold 5}} \\
Deep Tutor   & 325 & 256 & 33 & 103 \\
RoBERTa      & 341 & 256 & 17 & 103 \\
GPT-SE       & 322 & 210 & 36 & 149 \\
LLM Behavior & 357 & 5   & 1  & 2   \\
LLM No Def   & 357 & 7   & 1  & 0   \\
LLM Constr   & 289 & 4   & 69 & 3   \\ \bottomrule
\end{tabular}%
}
\caption{K Fold test set confusion matrices}
\label{tab:all-tp-fp-fn-tn}
\end{table}

\section{Discussion}



We placed the three evaluated semantic similarity approaches in the most favorable position by finding the optimal threshold using training sets of 5-fold cross validation. However, when compared to LLM-based approaches on the same dataset, all semantic similarity approaches performed worse than the two best-performing LLM-based approaches.
Our results suggest that LLMs could be an effective replacement for similarity-based methods for 2 reasons: they can operate without a ground-truth response and do not require a threshold to determine the correctness of student explanations. 
By defining rubrics, such as with Code Behavior Prompt, we achieved strong results with LLMs that are not state-of-the-art (like GPT-3.5) yet still beat contemporary semantic similarity methods.
The problem with using similarity-based approaches is that the choice of threshold plays a key role in the final prediction of student correctness.
Using nonlinear LLM approaches allows us to capture more information and features from the student explanations that are not categorized solely as lexical or semantic, but instead as latent-space representations that could have meaning or intent similar to that of the expert. 

\section{Conclusion}
In this work, we presented experiments that 
compared semantic similarity with LLM methods.
The LLM methods achieved top F1 and Accuracy scores without threshold tuning or fine-tuning, and better accommodating diverse student writing styles. 

%


\section{Limitations}
Our work is limited to assessing the LLM-powered assessment for student explanations. As future work, we could compare our methods across more datasets to support comparisons between semantic similarity and LLM-powered assessments, and to evaluate other student short-answer formats~\cite{li-etal-2025-chain-thought,liu-etal-2025-erevise}.

A limitation of the original dataset~\cite{selfcode2_dataset} is the variation among human experts in their assessments of student explanations. Discourse level~\cite{chen-etal-2025-threading} analyses could simplify and diversify the annotation of texts, which we will investigate further within the context of code explanations. 

We generated synthetic negative explanations using LLMs and briefly verified with 2 annotators using a sample of 100 explanations. For a more extensive evaluation we could compare synthetically generated and student generated explanations pedagogically by focusing on student misconceptions or gaps~\cite{oli2024can}. 

We did not employ systematic prompt optimization techniques such as those proposed by~\cite{dasgreater,khattab2023dspy}. Incorporating such methods — along with more clearly defined rubrics~\cite{zhou2022large,zhang2024reviseval}, which are effective — could further improve LLM-based assessment performance. Prompt optimization both for the generation and assessment of explanations would be an interesting extension of this work.

We did not use methods such as GRPO, recently explored for educational applications~\citep{duan2026kaser}. Preference alignment could be used to align LLM reasoning more closely with expert assessments and to generate synthetic incorrect explanations that better reflect authentic student misconceptions and fall within the distribution of typical student errors.

We did not explore using our LLM method to more programming languages, targeted at higher levels of student learning~\cite{biggs2014evaluating}. We also did not explore extensions to LLM-based assessment beyond correctness classification to generate targeted feedback for students.

\section*{Acknowledgements}
Thanks to Dr. Xiang Lorraine Li for her contributions and ideas!

\bibliography{custom}

\appendix

\section{Prompts}

\subsection{LLM-as-a-Judge}
Before finalizing the prompt, we manually reviewed 200 student explanations across the similarity range. We observed that some students focused on explaining the line's behavior (behavior-focused explanations). In contrast, others concentrated on why the line was used to achieve the program's goal (construction-focused explanations). To avoid unfairly assessing students who favored one of these explanation styles, we developed 3 versions of the LLM prompt, each with a different definition of correctness at the beginning. 
\begin{figure*}[h!]
    \centering
    \small
    \begin{mdframed}[align=center]

            Determine the correctness of the student's explanation based on a given source code, specific line number. \\
            
            (a) Set "Correctness" to 1 if the student's explanation is correct and 0 if it is incorrect.
            
            OR
            
            (b) "Correctness" A correct EXPLANATION explains why the line is used while implementing this program given the PROGRAM DESCRIPTION and SOURCE CODE below.
            
            OR
            
            (c)  "Correctness" A correct STUDENT EXPLANATION explains the code behavior using the code syntax. 
            
            ONLY set "Correctness". DO NOT PROVIDE REASON.
            
            PROGRAM DESCRIPTION 
            
            \{program desc\}
            
            SOURCE CODE 
            
            \{source code\}
            
            LINE NUMBER 
            
            \{line num\} 
            
            STUDENT EXPLANATION 
            
            \{student explanation\}
            
            CODE 
            
            \{line content\}
            
            Correctness:\\

    \end{mdframed}
         
        \caption{Prompting LLM to evaluate students' explanations. ``OR'' notation is used in this to indicate only one of the 3 possibilities were used to define ``Correctness'' to the model (a) = No Definition, (b) =  Construction, (c) = Code Behavior Prompt}
        \label{fig:prompt_llm_exp_eval}
\end{figure*}

\subsection{LLM for Generating Negative Examples}\label{sec:gen_neg_ex}
\begin{figure}[h!]  
\centering
\small
\subfloat[Roberta Similarity between Expert and Student Explanations in Original Dataset]
{
\includegraphics[width=\columnwidth]{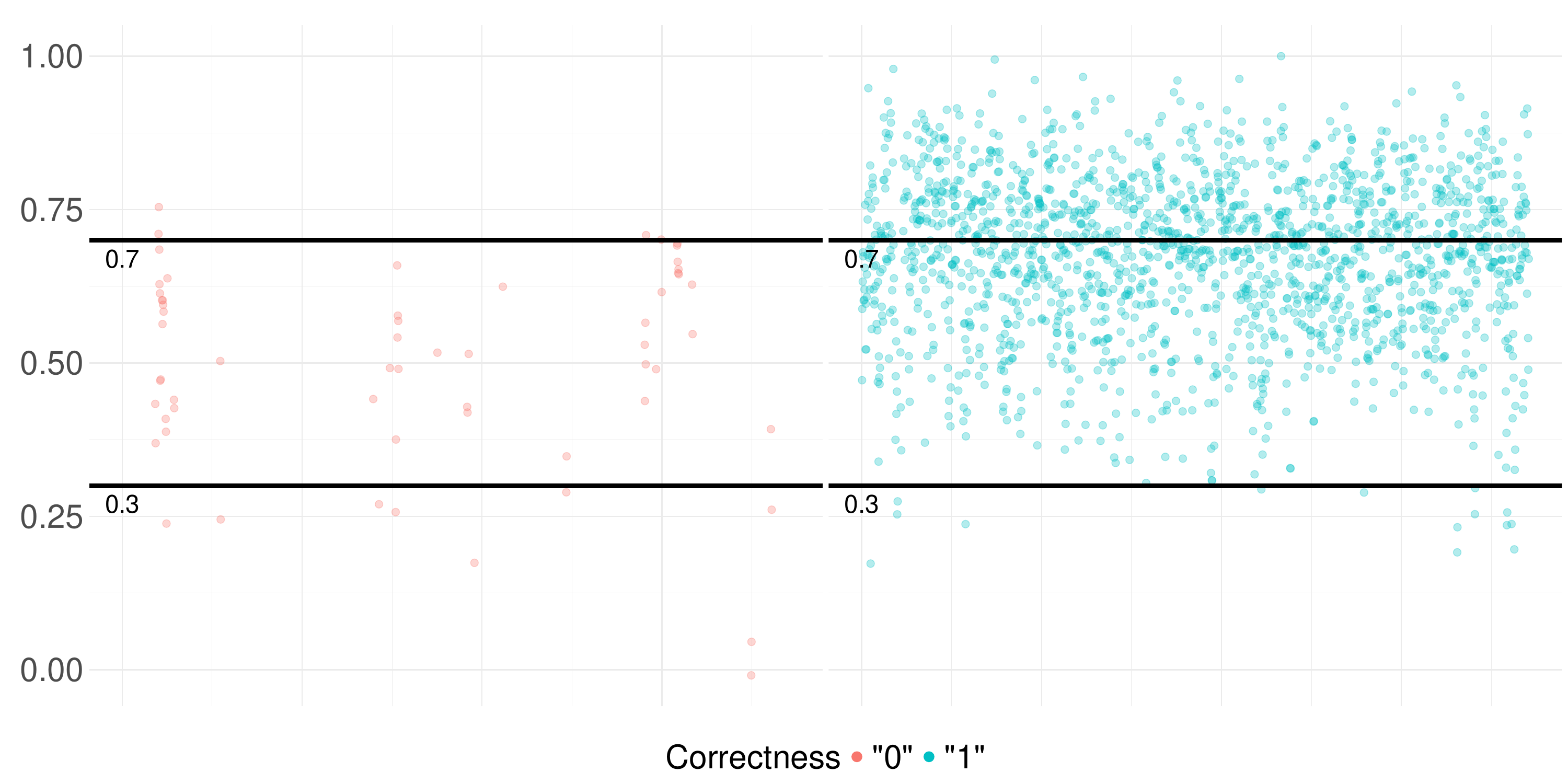}%
\label{fig:roberta_imbalanced}%
}
\subfloat[Roberta Similarity after Adding AI generated Explanations]
{\label{fig:roberta_balanced}%
\includegraphics[width=\columnwidth]{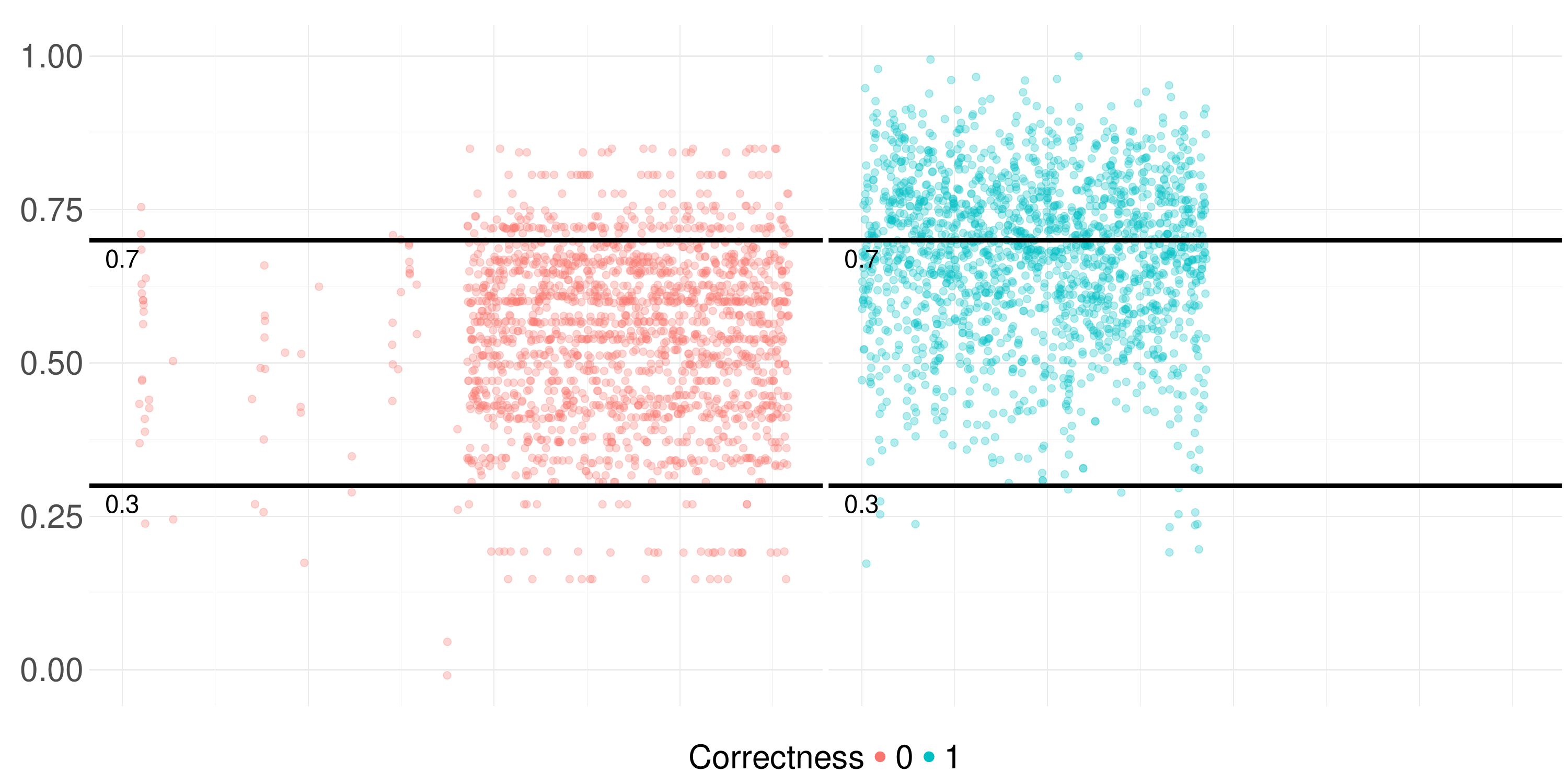}%
}
\caption{The AI-generated explanations remain similar to Student explanations, while also adding more negative examples (lower than 0.7, like student explanations) to balance the classes.}
\label{fig:semscore_distributions}
\end{figure}

To verify if the GPT-OSS 20b generated negative explanations that were indeed incorrect,  we sampled 100 explanations from the dataset and 2 researchers from the team independently annotated and confirmed that at least 95\% of the sample had incorrect explanations with an agreement of 91\%. The annotated samples from the 2 annotators are available for review on Drive~\footnote{\url{https://docs.google.com/spreadsheets/d/102WoJ1Y1EhtCIQcuWoGXav1HAkqsqMSE6a-ZdtfmxUw/edit?usp=sharing}}. One of the annotators further confirmed that an additional 100 explanations~\footnote{\url{https://drive.google.com/file/d/135O43Q2XexjrNG6XBriuEHjBnwe3eIA3/view?usp=sharing}} also had at least 95\% incorrect explanations.  We compared the distributions of semantic similarity scores for the generated negatives against those from the original dataset to confirm they preserved the character of authentic negative explanations (see Figure~\ref{fig:semscore_distributions}).

\begin{figure*}[h!]
    \centering
    \small
    \begin{mdframed}[align=center]
    \# Task  
Generate **3 Incorrect STUDENT EXPLANATIONS** for the specified **LINE NUMBER** in the given **SOURCE CODE**.

\#\#\# Correctness 

- The **STUDENT EXPLANATION** must accurately describe the behavior of the specified line of code. 
- It should correctly reference syntax, logic, and execution details of the **SOURCE CODE**.

---

\#\# Provided Information  

**Program Description:**  

{program\_desc}  

**Source Code:**  

{source\_code}  

**Line Number:**  

{line\_num}  

**Code:**  

{line\_content}  

**Incorrect Student Explanations:**  

{student\_explanation}  

**Expert Explanation:**  

{expert\_explanation}  

---
\#\# Output Format  

Return the final answer **only in valid JSON** with the following structure:

\{

  "incorrectExplanations": [
    "explanation 1",
    
    "explanation 2",
    
    "explanation 3"
  ]
  
\}
    
    \end{mdframed}
         
        \caption{Aritifical Generation to augment the original dataset with negative (incorrect) code examples }
        \label{fig:prompt_generate_negative_examples}
\end{figure*}

\end{document}